\documentclass[10pt,journal,compsoc]{IEEEtran}

\usepackage{url}
\usepackage{graphicx}
\usepackage{balance}
\usepackage{xcolor}
\usepackage{fancyhdr}

\fancypagestyle{firststyle}
{
   \fancyhead[C]{\LARGE{\textbf{To appear in IEEE Software}}}
}

\begin{document}

\title{The Four Pillars of Research Software Engineering}

\author{
	\IEEEauthorblockN{Jeremy Cohen\IEEEauthorrefmark{1}, Daniel S. Katz\IEEEauthorrefmark{6}, Michelle Barker\IEEEauthorrefmark{3}, Neil Chue Hong\IEEEauthorrefmark{4}, Robert Haines\IEEEauthorrefmark{5}, Caroline Jay\IEEEauthorrefmark{5}}\\
	\vspace{1em}
	\IEEEauthorblockA{\IEEEauthorrefmark{1}Department of Computing, Imperial College London, London, UK. \emph{jeremy.cohen@imperial.ac.uk}}\\
	\IEEEauthorblockA{\IEEEauthorrefmark{6}University of Illinois at Urbana-Champaign, Urbana, IL, USA. \emph{d.katz@ieee.org}}\\
	\IEEEauthorblockA{\IEEEauthorrefmark{3}Australian Research Data Commons, Melbourne, Australia. \emph{michelle.barker@ardc.edu.au}}\\
	\IEEEauthorblockA{\IEEEauthorrefmark{4}Software Sustainability Institute, University of Edinburgh, Edinburgh, UK. \emph{n.chuehong@software.ac.uk}}\\
	\IEEEauthorblockA{\IEEEauthorrefmark{5}University of Manchester, Manchester, UK \emph{\{robert.haines, caroline.jay\}@manchester.ac.uk}}
}

\maketitle

\thispagestyle{firststyle}

\abstract Building software that can support the huge growth in data and computation required by modern research needs individuals with increasingly specialist skill sets that take time to develop and maintain. The Research Software Engineering movement, which started in the UK and has been built up over recent years, aims to recognise and support these individuals. Why does research software matter to professional software development practitioners outside the research community? Research software can have great impact on the wider world and recent progress means the area can now be considered as a more realistic option for a professional software development career. In this article we present a structure, along with supporting evidence of real-world activities, that defines four elements that we believe are key to providing comprehensive and sustainable support for Research Software Engineering. We also highlight ways that the wider developer community can learn from, and engage with, these activities.

\section{Introduction}

While software has been a part of research for many decades, the people who write, maintain and manage this \textit{research software} are increasingly seen as critically important members of research teams, rather than just ``the people who write code''. The past few years have seen the emergence and rapid growth of the concept of \textit{Research Software Engineering} (RSE). Starting out in the UK research community, from around 2013, through discussions initiated by the Software Sustainability Institute (\url{https://software.ac.uk}), groups  focusing on developing sustainable, maintainable, robust research software have been set up at many institutions. A history of this process can be found in the 2017 Research Software Engineers: State of the Nation report~\cite{rse-sotn}. Since the report was written, the number of research software groups within the UK and internationally has increased. Behind this growth is the need for ever-increasing amounts of high quality research software development. If you are a software development practitioner, addressing this ultimately relies on people like \textit{you} seeing research institutions as a potential employer. For researchers building software, it relies on seeing software development as a realistic focus for a research career. While the majority of Research Software Engineers (RSEs) currently come from the research community and there is not yet a well-established path for professional developers moving into the field, a small number of RSEs already come from a professional development background. RSE-specific roles and job descriptions are making this a realistic career option and anecdotal evidence suggests that the number of professional developers moving into RSE is growing. Although there can be similarities between RSE roles and those in major tech companies or the startup community, Cannam, et al.'s ``Ten reasons to be a research software engineer''~\cite{ref:ten-reasons} provides some great examples of why a professional developer might choose an RSE career over alternative options.

Ensuring a sustainable approach to research software development isn't just about how you build the software, but also how you support the community of people building it, and how you attract the most talented individuals, both professional developers and researchers, to join this community. In this article we introduce ``4 pillars'' of Research Software Engineering that we see as providing the necessary elements to offer comprehensive, ongoing support for the development of quality research software. The structure aims to demonstrate how professional software development practices can be brought together with approaches used in the research community to provide a more sustainable environment for developers of research software. The basis for, and structure of, the 4 pillars were initially set out in our earlier short paper~\cite{wssspe-pillars} as a request for comments and input from the community. This article supports the previously proposed structure and provides a range of evidence to show how the areas represented by the 4 pillars are being realised. We show that these areas include extensive professional software development practitioner knowledge that is now more likely to be recognised as important and necessary in the development of robust research software. Adoption of elements of the structure is already leading to research rapidly becoming a viable environment for a professional software engineering career with the benefits that come with working in a flexible environment characterised by interesting research challenges.

This article aims to provide insights in three core areas:
\begin{itemize}

\item Understanding the importance of Research Software Engineering to the research community and to the wider developer community and why the field has grown so rapidly in recent years.

\item Setting out a structure that defines the core elements of Research Software Engineering. This structure can support an organisation in ensuring reliable, maintainable, and sustainable research software outputs and reproducible research.

\item Examples of existing activities that demonstrate the areas and approaches highlighted by the 4 pillars.

\end{itemize}

\section{The importance of research software}

Software increasingly underpins almost all research and, indeed, industry and much of the wider economy. Since the earliest days of using software to support and undertake research, there have been links with industry. However, with the emergence of the data economy, enterprises increasingly need to adopt advanced computational processes and pipelines for managing and understanding the vast quantities of data available to them. The speed of development in areas such as data science, machine learning, and artificial intelligence creates a much more direct link between research software outputs, industry and the wider world. There are some excellent examples of how open-source software developed to support the research community is being used to tackle challenges in the wider world, for example, the use of Jupyter notebooks for large-scale data analysis by Netflix~\cite{ref:netflix-notebooks}. Research funding has also been behind some key developments in modern society, for example, a research project at Stanford that lies behind the development of the Google search engine~\cite{ref:origins-google}. Indeed, research projects and funding have helped to support the development of what could be considered to be some of the most fundamental and life-changing aspects of the modern world, including the Internet itself, and the protocols that power it~\cite{ref:internet-history}.

Ultimately, good research software can make the difference between valid, sustainable, reproducible research outputs and short-lived, potentially unreliable or erroneous outputs. This is succinctly highlighted by the tagline of the UK Software Sustainability Institute: ``\textit{better software, better research}''. Good research software is difficult and time consuming to build, but it matters. Despite this, as pointed out by Anna Nowogrodzki's ``How to support open-source software and stay sane''~\cite{ref:nowogrodzki19}, funding the development of research software can be very challenging.

\begin{figure*}[ht]
   \centering
   \includegraphics[width=14cm]{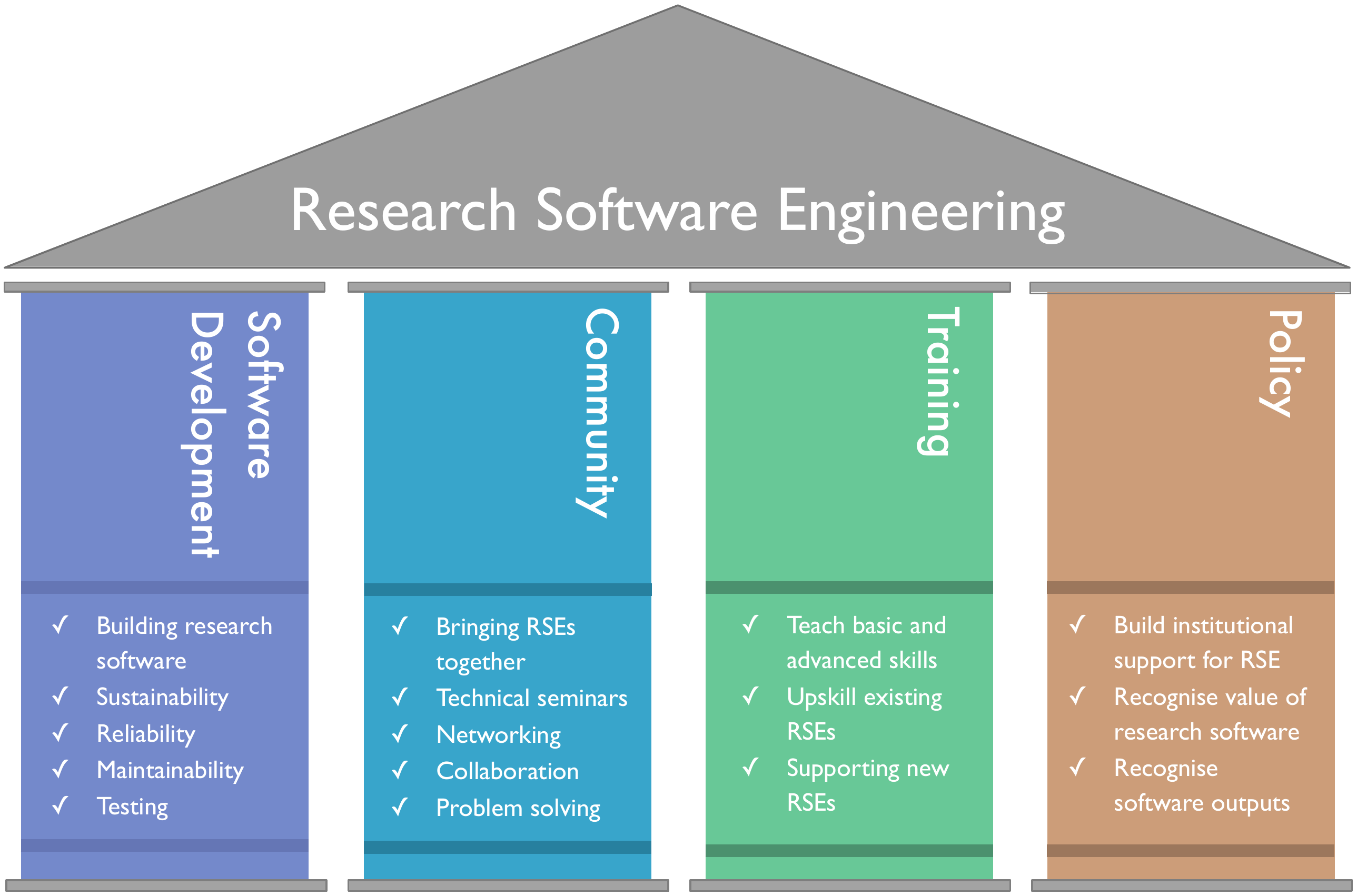}
   \caption{The 4 pillars of Research Software Engineering}
   \label{fig:4pillars}
\end{figure*}

\section{4 pillars of research software}
\label{section:pillars}

Our observation is that organisations' support for research software often develops in an ad hoc manner. Our 4-pillar structure, summarised in this section, aims to highlight areas that we see as most important in providing comprehensive support for research software. Research software is built in a wide range of different organisations from large universities to small independent research labs. The scale and variety of research software-related activities will differ between organisation types. The 4-pillar structure aims to offer the complete picture, however, there will be some organisations for which a subset of the activities described is sufficient. We are not suggesting that organisations that only provide support for, say, the software development pillar, are worse places to be a software developer. The structure offers organisations a straightforward approach to identifying where current activities could be extended. For existing research software practitioners looking to move into this space, it provides a way of gauging how advanced an organisation is in this area. We hope it begins the process of helping to formalise the space within which research software is built.

\subsection{When the coding gets tough}

When we talk about ``better'' software or ``good quality'' software, what do we actually mean in the research context, and why is it important? Many researchers who write software are, to some extent, self-taught. They may have had some basic software development training as part of an undergraduate or postgraduate course but the first time they actually write a tool, script, or application to serve some real-world requirement is when they hit a research problem that can best be solved with some code. At this stage, getting the end result is often the most important aspect, and it is often needed quickly.

Best practices covering documentation, testing, software design techniques, and code management are key here. They help to ensure that software is reliable, easier to maintain, and that the results it outputs are reproducible. While the Research Software Engineering movement can provide training in these skills to researchers, they are a natural part of everyday software development for professional software practitioners who can contribute significantly here.

Maintainability, sustainability and robustness are core aspects of building quality software that form part of our first pillar -- \textbf{Software Development}.

\subsection{No developer is an island}

Software is an incredibly fast-changing field. New frameworks and tools can appear, gain huge traction in the community with thousands or even millions of users, and then a couple of years later be almost forgotten and replaced with the ``next big thing''. How do developers find out about such changes and, more importantly, how do they know which of these changes are relevant, which should be learnt/adopted, and which should be avoided. What common technical challenges do they face?

In the modern world of software, it is very difficult for developers to work entirely alone. However, someone building research software within an academic research group can frequently be the only developer within their team, or one of a small group of such people. Life as a research developer can sometimes be lonely! RSEs in central research software engineering teams may have peers to communicate with, but interacting with other developers from different fields and technical backgrounds can always be beneficial. Professional software practitioners can also bring a lot to this space in the form of technical advice and guidance.

The need for communication, learning from the experiences of others, and keeping up with the latest developments provides the basis for our second pillar -- \textbf{Community}.

\subsection{New code on the block}

As a developer, there's always something new to learn, partly because of the speed of change in the developer community (and computing in general) and partly because it is such a large area with so many different languages, techniques, and tools. In addition to keeping coding skills up-to-date, developers also benefit from guidance on tools and approaches for making software open, citable, robust, readable, verifiable, and easier to reuse and contribute to. 

The concept of Continuing Professional Development (CPD) is now commonplace in many industries but is less common in the research community, perhaps because the whole process of research is based around learning/developing new skills and techniques. In our experience, research software developers are enthusiastic about training opportunities. 

The importance of ensuring initial and ongoing skills development for people building research software underpins our third pillar -- \textbf{Training}.

\subsection{Change the world}

Ensuring good software development practices and community and training support for developers is important, but in our opinion, it doesn't cover the full picture required by a comprehensive supporting environment for research software. The missing element is the need to develop and provide strong institutional processes that recognise the importance of developing quality software as a key to strengthening research outputs. These processes should also include support for research software careers and cover both researchers with a focus on developing software and software practitioners whose main task is supporting researchers. There is also a need for higher level support provided through government policy and funding approaches. 

Developing both institutional and national policies that recognise the importance of research software provides the basis for the fourth pillar -- \textbf{Policy}.

While all 4 pillars contribute to the cultural change required, changes to policy frameworks can provide significant top-down impact to support the other 3 pillars. Positioning policy within the framework is, therefore, challenging since the other pillars can be seen as relying on, and building on, policy. However, the organic growth of research software activities at institutions means that we observe many cases where other activities are already in place before policy aspects have been considered or addressed. We therefore include policy within our framework as a pillar supporting the overall RSE space, alongside the other pillars.

\vspace{1em}
\noindent Figure~\ref{fig:4pillars} summarises the 4 pillars. These 4 pillars are similar to the elements of social change models that focus on individual, organisation, community, and policy levels. One example is Nosek's strategy for cultural change (utilised by the Centre for Open Science), which identifies five levels -- infrastructure, user interface/experience, communities, incentives and policy~\cite{ref:nosek19}. 

\section{RSE in the wild}

We now look at some examples of how aspects of our 4-pillar structure are being realised ``in the wild'' by different organisations around the world.

\noindent \textbf{Software Development:} It is probably fair to say that software development methodologies familiar to the professional software developer community have traditionally not been applied in an environment where one or two self-taught research developers are building some software to solve a research challenge. Having said that, the approaches these developers often end up using actually have similarities to modern agile or rapid application development methodologies. The process of building a quick prototype, getting feedback and then iteratively updating and reviewing the code is very useful in a research environment. This is similar to the ``Release Early, Release Often'' principle described by Eric S. Raymond~\cite{ref:raymond00} in the context of the development of the Linux kernel. While perhaps not intuitively conducive to ensuring the release of quality, reliable software, this is now recognised as a valuable approach in many scenarios. For open-source software, putting code where everyone can see and comment on it can help to gain valuable feedback and even develop new collaborations that can lead to better tested, better quality outputs. 

Katz et al. have looked at case studies of developing research software at three institutions in the US and UK, focusing on how their staff are organised and the models and processes that they use~\cite{ref:katz19}. In some cases, research software groups undertake activities that go beyond the software development pillar, e.g. training or running communities, while in others, they focus purely on software development.

Over the last couple of years, the UK's Software Sustainability Institute, in collaboration with partners in several other countries, has undertaken an international survey of Research Software Engineers. The most recent survey, from 2018~\cite{ref:rse-survey2018}, covers RSEs in 8 countries and demonstrates how the use of software best practices in areas such as testing is improving over time.

\noindent \textbf{Community:} Research software communities can exist at four different levels -- local, regional, national and international. 

Local communities generally exist within an institution and support researchers and RSEs with the software development aspects of their roles and networking with peers within their institution. These communities aim to raise the profile of software, and of best practices for developing it. Examples of local communities include Imperial College London's research software community~\cite{ref:imperial-rs-community} and the software engineering community at DLR, the German Aerospace Center~\cite{ref:dlr-community}. The example of DLR also highlights another general class of RSE communities that are discipline specific. In addition to bringing developers and researchers together, community members can become aware of and engage more effectively with open source software being developed across a community, driven by the release early and often paradigm highlighted above.

National communities serve a more strategic role in coordinating national research software activities and advocating for better recognition and support for research software professionals. Communities such as the DevLOG network (http://devlog.cnrs.fr/) in France have existed for some time, providing support and skills development opportunities to individuals undertaking software-related work. Following the coining of the ``RSE'' name and the development of the UK RSE community, national communities have now been set up in other countries/regions. These include Germany (deRSE -- \url{https://www.de-rse.org}), the Netherlands (NL-RSE -- \url{https://nl-rse.org/}), the Nordic region (Nordic-RSE -- \url{http://nordic-rse.org/}), the US (US-RSE -- \url{https://us-rse.org/}) and Australia/New Zealand (RSE AU/NZ community -- \url{https://rse-aunz.github.io/}). While we accept that this is a small number of countries, we see scope and interest in other countries and believe this is a movement that will continue to expand. The most recent development has been the creation of a professional society for RSEs: the Society of Research Software Engineering (https://society-rse.org/) in the UK.

We are beginning to see the emergence of regional research software communities that bridge the gap between local and national communities. An example is the Research Software London community (\url{https://rslondon.ac.uk}) for London and the South East of England. Regional communities provide opportunities to learn from the different approaches taken at geographically local institutions, to share the organisation of community activities, and to support and engage smaller institutions that may not have a critical mass of people to set up and sustain their own local community.

An international community level is also emerging with the formation of Research Software Engineers International in 2018, after leaders of national RSE associations, groups and related initiatives from around the world came together in London for the first International RSE Leaders Workshop, and with the launch of the Research Software Alliance (\url{https://www.researchsoft.org/}). Other international communities such as WSSSPE -- Working Towards Sustainable Software for Science: Practice and Experiences (\url{http://wssspe.researchcomputing.org.uk}) -- bridge the core research software community and more research-focused aspects of software engineering and computing covering areas such as empirical software engineering.

\noindent \textbf{Training:} Research and technology are on a never-ending progression of advancement and change. Software is part of this. Training is therefore vital to ensure that the value of research software is understood. It should focus not only on the purely technical skills of writing code, but also on how to apply these skills in a manner that results in sustainable, reusable and maintainable software.

There is a diverse range of materials and approaches for training. The Carpentries (\url{https://carpentries.org/}) provide an important set of base-level training material covering software, data, library/information tools and high performance computing. CodeRefinery provide a series of carpentry-style lessons~\cite{ref:coderefinery} that include some more advanced technical material but also cover best practices around how software can support research through aspects such as reproducibility. Many institutions also provide their own in-house training courses, which are sometimes delivered by members of a central RSE team. For example, RSE groups in US, UK and German universities and research organisations are increasingly becoming the hub for digital training at their institutions. Such groups are helping to play a key role in enabling sustainable and scalable delivery of the Carpentries to the research community. There are also now an increasingly large number of meetup-style groups that offer a more informal approach to training, through technical seminars, for example. These groups are often not tied to a specific organisation, and their ability to attract people from a wide range of academic institutions and industry can provide a great opportunity to encourage and support diversity. An example is the international R-Ladies community (\url{http://rladies.org}). 

\noindent \textbf{Policy:} There are now many active groups helping to develop, support, and influence policy to aid the research software community. An issue of particular importance and something that is frequently raised at ``grass roots'' gatherings of RSEs is careers. This is important not only for individuals already in a research software role, but also for attracting software practitioners from other fields into research. 

Providing roles and career paths tailored to individuals supporting research is, of course, not new. The ``\textit{Research Engineer}'' role that is common in the French research community is one example. It encompasses individuals undertaking a wide range of research-related activities, of which software development is one key area. In countries where there has not been a direct counterpart to this role for individuals focusing on research software, the RSE movement has provided a way to recognise, support and offer career paths to developers of research software. As their numbers have grown, the challenge that they face in finding a long-term, sustainable career path has also become more obvious.

A recent paper on RSEs in the UK provides a good summary of some of the types of policy change needed, detailing the role that employers and funding bodies can play in career progression~\cite{ref:importance-softres}. This paper also points out the crucial role of measuring the impact of RSE groups in improving the efficiency of research projects and, in turn, saving time and money and improving the quality of research outputs. This kind of data can influence government policy to recognise the importance of both software and those who develop it. More of this evidence is needed. 

Research software teams benefit strongly from having a combination of software practitioners from both industry and research backgrounds. However, research credit is heavily biased towards traditional research outputs such as papers. This puts off many researchers from considering a research software career. While it has been possible to obtain unique identifiers, Digital Object Identifiers (DOIs), for papers for quite some time, obtaining DOIs for software and data, and having accepted and agreed means to cite these research assets is comparatively recent. It is important that research software is recognised as a first-order element of research. Policy makers have a very strong role to play here and things are now changing in this area. The FORCE11 Software Citation Working Group created a set of software citation principles~\cite{ref:citation-principles}. Set up in 2013, Zenodo (\url{https://zenodo.org}) provides a means to obtain DOIs for software and data assets. In collaboration with Github, details are provided on how to obtain a DOI to cite a GitHub repository~\cite{ref:github-citation}. It is not just the need to be able to cite software that is important. Di Cosmo and Zacchiroli, in their description of the Software Heritage initiative~\cite{ref:software-heritage}, provide a strong case for the need to preserve the information represented by, and within, source code and their work to address this.

\section{Sustaining and growing research software support}

In this article, we have presented 4 pillars representing a set of areas and activities that we consider to be vital in offering coordinated and comprehensive support for research software and the people who build it. In turn, we hope this will demonstrate to professional developers and researchers alike that research is a viable, and interesting, environment for a software development career. The wide-ranging need for large-scale data processing and computation in industry means that many companies now have technical roles that are similar to RSE roles. These can bring with them salaries and benefits with which research institutions may be unable to compete directly, but there are still reasons to choose an RSE career. The results of the previously mentioned 2018 International RSE Survey~\cite{ref:rse-survey2018} include ``Desire to advance research'' and ``Freedom to choose own working practices'' as some of the most highly-ranked responses to a question about reasons for choosing your current job. The career choices that professional developers have, and the demand for and value of their skills, makes it even more important for organisations to invest in the activities represented under the 4 pillars.

While we are currently unaware of any research institution/organisation that comprehensively implements all the aspects of the 4-pillar structure some institutions are close to achieving this and are already realising benefits from their support of research software activities. However, as highlighted in Section~\ref{section:pillars}, we recognise that full realisation of this structure within an organisation, where some form of support is provided for each of the 4 pillars, may not always be practical, realistic, or even necessary. We have tried to present examples of existing activities under the different pillars to give an idea of how the various elements of our structure may be realised and to demonstrate to the reader how they might be able to help to support the development of more effective, sustainable, reliable research software. Providing this support is important in ensuring the quality and longevity of research outputs and in attracting professional software practitioners to the research community.

\section{Acknowledgements}

The authors would like to thank a number of individuals who have initiated and supported the development of RSE leading to the activities used as examples in this article. While it is impractical to name everyone involved, we would particularly like to acknowledge James Hetherington, Simon Hettrick and Mike Croucher who have been at the forefront of developing RSE groups and supporting infrastructure in the UK. JC acknowledges UKRI EPSRC for support from grant EP/R025460/1. DSK acknowledges support from NSF 1743188. CJ/NCH acknowledge support from EPSRC EP/S021779/1 and NCH acknowledges EP/N006410/1.

\bibliographystyle{unsrt}
\balance
\bibliography{4PillarsOfRSE}

\end{document}